\documentclass[10pt,reprint,twocolumn,prd,aps,amsmath,amssymb,nofootinbib]{revtex4-1}

\usepackage{amsfonts}
\usepackage{textcomp}
\usepackage{gensymb}
\usepackage{graphicx}
\usepackage{needspace}
\usepackage[colorlinks=false]{hyperref}

\allowdisplaybreaks

\begin{document}

\title{Perturbing Monopolar Force-Free Magnetospheres to Slowly Rotating Black Holes}

\author{Kevin Thoelecke}
\affiliation{Department of Physics, Montana State University, Bozeman, Montana 59717, USA}

\author{Masaaki Takahashi}
\affiliation{Department of Physics and Astronomy, Aichi University of Education, Kariya, Aichi 448-8542, Japan}

\author{Sachiko Tsuruta}
\affiliation{Department of Physics, Montana State University, Bozeman, Montana 59717, USA}

\begin{abstract}
We study the effects of frame-dragging on the structure of force-free magnetospheres around rotating black holes.  For slowly rotating black holes, analytic explorations often focus on transforming an exact force-free solution applicable to a static black hole into a solution approximately applicable to a slowly-rotating black hole via perturbations in black hole spin.  We show that the single perturbed monopolar solution most commonly arrived at using such techniques is in fact a separatrix between two broad classes of solutions: those with poloidal magnetic fields that bend upwards towards the azimuthal axis and those that bend downwards towards the equatorial plane.  This is because frame-dragging affects the toroidal magnetic field distributions, changing the force balance between poloidal magnetic field lines.
\end{abstract}

\maketitle

\section{Introduction}

In \citet{BZ77} a mathematical procedure was developed to perturb force-free magnetospheres from non-rotating to rotating spacetimes.  Using that procedure it was shown that a monopolar magnetic field around a slowly rotating black hole should rigidly rotate at roughly half the rate at which the horizon rotates, demonstrating that it is possible to efficiently extract a black hole's rotational energy via an outgoing Poynting flux.  The magnetosphere parameters associated with that monopolar solution are often used today to describe and estimate the expected physical attributes of an energy-extracting black hole magnetosphere.

What is sometimes missed is that the single monopolar solution and associated magnetosphere parameters found by \citet{BZ77} (and extended to higher order in later works such as  \citet{MG2004}, \citet{TanabeNagataki2008}, and \citet{PanYu2015}) is based on a special case of a broader class of monopolar magnetospheres applicable to flat spacetimes found by \citet{Michel1973} (that are themselves a subset of a broader solution space).  Specifically, the solution found by \citet{BZ77} is the result of perturbing around an initially non-rotating magnetosphere, while the solutions found by \citet{Michel1973} allow for arbitrary uniform rotation (non-uniform rotation is also conditionally allowed). 

From an analytic computational perspective there is a very practical reason for selecting an initially non-rotating magnetosphere; the perturbation techniques used are primarily useful when changes to the structure of the poloidal field are small \citep{BZ77}.  If the changes are not small then the non-linearity of the equations involved can lead to large corrections that reduce the utility of a perturbative approach.  The non-linearity in the equations is primarily sourced by field line rotation, so assuming vanishing initial rotation diminishes the probability of the emergence of problematic effects (at least in first order corrections).

In this work we more generally perturb the monopolar magnetospheres of \citet{Michel1973} in a manner compatible with the approach of \citet{BZ77} without demanding that the initial magnetosphere be non-rotating.  The non-linear nature of the equations necessarily limits the obtained solutions' regions of validity, and should a specific rotational profile be desired different analytic (or numeric) techniques from the somewhat ad hoc ones we employ would likely be more profitable.  Nonetheless the solutions obtained are sufficient to demonstrate that more slowly rotating magnetospheres (referenced to black hole spin) should be expected to have poloidal magnetic field lines that bend upwards towards the azimuthal axis, while more rapidly rotating magnetospheres should have poloidal magnetic field lines that bend downwards towards the equatorial plane (a result compatible with the numerical conclusions of \citet{TTT2017}).  The single monopolar (or split-monopolar) solution found by \citet{BZ77} and others is a separatrix between those two behaviors, implying much smaller changes to the structure of the poloidal magnetic field and therefore more compatibility with a perturbative approach.  

From a mathematical point of view this suggests that higher-order perturbative explorations of the common monopolar solution might become problematic if deviations from the separatrix solution emerge, potentially leading to apparent inconsistencies with a perturbative approach.  Should divergences or other poor behaviors emerge, they might be understood and corrected by considering how the solution deviates from the separatrix.  The separatrix solution might also be more usefully mathematically interpreted as a rotation-driven selection from a group of existing solutions instead of the reaction of a single solution to the addition of spacetime rotation. 

From a physical point of view this means that the landscape of energy-extracting black hole magnetospheres is much broader than the single monopolar solution found by \citet{BZ77} and others might suggest, and by extension that different rates of black hole energy and angular momentum extraction might be coupled to global magnetosphere structure.  That broader solution space allows for more flexibility in considerations of astrophysical scenarios in which black hole energy extraction might be relevant.  Instead of a single solution with a fixed rate of energy extraction for a given black hole spin, the rate of energy extraction may be tuned in a way that is coupled to magnetosphere structure.

We begin our analysis of the problem in Section \ref{Sec:BasicEquations} by stating the general equations and assumptions in use.  We then expand and more generally extend the monopolar magnetospheres of \citet{Michel1973} to Schwarzschild spacetimes in Section \ref{Sec:SchwarzschildMonopole}.  In Section \ref{Sec:KerrMostMonopolar} we arrive at the single extension of those magnetospheres found by \citet{BZ77} from a slightly different perspective.  We then solve for a more general case in Section \ref{Sec:KerrGenericExtension} before discussing behaviors and error of the solution in Section \ref{Sec:ErrorAnalysis} and concluding.



\section{Basic Equations and Assumptions} \label{Sec:BasicEquations}

We first assume a stationary and axisymmetric spacetime, expressed in Boyer-Lindquist coordinates as:
\begin{align}
ds^2 &= \left(1 - \frac{2mr}{\Sigma} \right) dt^2 + \frac{4mar \sin^2 \theta}{\Sigma} dt d\phi - \frac{\Sigma}{\Delta} dr^2 \nonumber \\
&- \Sigma d\theta^2 - \frac{A \sin^2 \theta}{\Sigma} d\phi^2,
\end{align}
where $m$ is the mass and $a$ is the angular momentum per unit mass of the black hole, and:
\begin{align}
\Sigma &= r^2 + a^2 \cos^2 \theta ,\nonumber \\
\Delta &= r^2 - 2mr + a^2, 	\nonumber \\
A &= \left(r^2 + a^2 \right)^2 - \Delta a^2 \sin^2 \theta .
\end{align}
We then assume a stationary and axisymmetric force-free magnetosphere, such that:
\begin{equation}
T_\alpha{}^\beta{}_{; \beta} = -F_{\alpha \beta} J^\beta = X_\alpha = 0.
\end{equation}
Here $T^{\alpha \beta}$ is the stress energy tensor, $F^{\alpha \beta}$ is the electromagnetic field strength tensor, $J^\alpha$ is the current vector, and $X^\alpha$ is the momentum flux vector.  The condition that $X^\alpha = 0$ may be expanded to find (\citet{TTT2017}):  
\begin{align} \label{Eq:ForceFreeCondition}
4 \pi \Sigma \sin \theta \frac{X_A}{A_{\phi, A}} &= - \frac{1}{2} \frac{\Sigma}{\Delta \sin \theta} \frac{d}{d A_\phi} \left(\sqrt{-g} F^{\theta r} \right)^2 \nonumber \\
&- \frac{1}{\sin \theta} \left(\alpha F_{r \phi} \right)_{, r} - \frac{1}{\Delta} \left( \frac{\alpha}{\sin \theta} F_{\theta \phi}\right)_{, \theta} \nonumber \\
&+ \frac{G_\phi}{\sin \theta} \left(F_{r \phi} \Omega_{\text{F}, r} + \frac{1}{\Delta} F_{\theta \phi} \Omega_{\text{F}, \theta} \right),
\end{align}
where:
\begin{align}
\alpha &= g_{tt} + 2 g_{t \phi} \Omega_\text{F} + g_{\phi \phi} \Omega_\text{F}^2, \nonumber \\
G_\phi &= g_{t \phi} + g_{\phi \phi} \Omega_\text{F}.
\end{align}
Here the uppercase Latin indices ($A$) indicate poloidal $(r, \theta)$ directions and a comma denotes a partial derivative.  Stationarity, axisymmetry, and the assumption of a perfectly conducting plasma (expressed as the vanishing contraction of the electromagnetic field strength tensor with its dual) allow us define the field line angular velocity $\Omega_\text{F}$ as:   
\begin{align} \label{Eq:OmegaFDefinition}
F_{tr} &= F_{r \phi} \Omega_\text{F}, \nonumber \\
F_{t \theta} &= F_{\theta \phi} \Omega_\text{F}. 
\end{align}
This is a statement of the rigid rotation of magnetic field lines; the toroidal vector potential $A_\phi$ (in our coordinate basis $F_{\alpha \beta} = A_{\beta, \alpha} - A_{\alpha, \beta}$) traces poloidal magnetic field lines, and under the above conditions surfaces of constant $A_\phi$ coincide with surfaces of constant $\Omega_\text{F}$. 

Stationarity and axisymmetry demand conserved fluxes of energy $E$ and angular momentum $L$ such that $E = E(A_\phi) = (1/4\pi)\sqrt{-g} F^{\theta r} \Omega_\text{F}$ and $L = L(A_\phi) = (1/4\pi) \sqrt{-g} F^{\theta r}$; this in turn demands that the toroidal magnetic field $\sqrt{-g} F^{\theta r}$ and magnetic field line angular velocity $\Omega_\text{F}$ be conserved (see, eg., \citet{BZ77}):
\begin{align}
\sqrt{-g} F^{\theta r} &= \sqrt{-g} F^{\theta r} (A_\phi), \nonumber \\
\Omega_\text{F} &= \Omega_\text{F}(A_\phi).
\end{align}
If those conditions are not met then $X_t \neq 0$ and/or $X_\phi \neq 0$ and satisfying $X_A = 0$ in Equation \ref{Eq:ForceFreeCondition} would not yield a self-consistent $X^\alpha = 0$ solution.  

Equation \ref{Eq:ForceFreeCondition} is insensitive to the sign of the toroidal field, which is to say insensitive to the inward/outward direction of Poynting flux.  If an ingoing observer on the horizon is to measure finite electromagnetic fields, however, then the toroidal field must satisfy the Znajek regularity condition on the horizon (\citet{Znajek1977}):
\begin{equation} \label{Eq:ZnajekCondition}
\sqrt{-g} F^{\theta r} \left(r_\text{H}, A_\phi\right) = - \frac{\left(r_\text{H}^2 + a^2 \right) \left(\Omega_\text{F} - \omega_\text{H} \right) \sin \theta_\text{H}}{\Sigma_\text{H}} A_{\phi, \theta}^\text{H}.
\end{equation}
Here $\omega_\text{H} = a / 2 m r_\text{H}$ is the angular velocity of the horizon.  This regularity condition does not restrict the solution space in any significant way, as its square is already present in Equation \ref{Eq:ForceFreeCondition}; the only additional information it offers is which solution for the toroidal field (ingoing or outgoing Poynting flux) is physically valid on the horizon.  Nonetheless it can still be a very useful simplification of Equation \ref{Eq:ForceFreeCondition} when the horizon is being considered. 


\section{Schwarzschild Monopole Solution} \label{Sec:SchwarzschildMonopole}
In this section we arrive, in abbreviated form, at the monopolar solution found by \citet{Michel1973} in the context of flat space and extended by \citet{BZ77} to slowly rotating spacetimes.  We refer to it as a ``monopolar'' solution because the magnetic field in the poloidal plane is monopolar, but there can be a toroidal component of the magnetic field arising from magnetosphere rotation.\footnote{Often such solutions are referred to as ``split-monopolar'' by adding reflection asymmetry across the equatorial plane (in order to make the magnetic field divergenceless when integrated over a volume enclosing the black hole); such distinctions are irrelevant for our current purposes, however, and can trivially be addressed by applying the restriction $0 \leq \theta \leq \pi/2$ everywhere.}  A monopolar poloidal magnetic field around a Schwarzschild black hole may be described by the toroidal component of a vector potential given as:
\begin{equation}
A_\phi = B_0 \cos \theta.
\end{equation}
This yields $F_{\theta \phi} = -B_0 \sin \theta$ and a magnetic field far from the black hole that is given by $\mathbf{B} \sim (B_0/r^2) \hat{r}$ in standard orthonormal spherical coordinates.  For such a field to be force-free everywhere, far from the black hole we demand that (Equation \ref{Eq:ForceFreeCondition} as $r \rightarrow \infty$):
\begin{equation} \label{Eq:InfinityCondition}
\left[\left(\sqrt{-g} F^{\theta r} \right)^2\right]' = B_0^2 \sin^4 \theta \left(\Omega_\text{F}^2 \right)' - 4 B_0 \cos \theta \sin^2 \theta \Omega_\text{F}^2.
\end{equation}   
Here a prime denotes a derivative with respect to $A_\phi$.  Therefore the field line angular velocity as $r \rightarrow \infty$ can be specified as an arbitrary function of poloidal angle.  Insertion of this condition into Equation \ref{Eq:ForceFreeCondition} shows that it holds for all radii, including the horizon (as may be intuited from the coincident symmetry of the spacetime with the electromagnetic fields), yielding a general solution that is given by:  
\begin{align} \label{Eq:SchwarzschildMonopoleSolution}
A_\phi &= B_0 \cos \theta, \nonumber \\
\Omega_\text{F} &= \Omega(\theta), \nonumber \\
\sqrt{-g} F^{\theta r} &= B_0 \Omega(\theta) \sin^2 \theta.
\end{align}
To arrive at the above we have applied the Znajek regularity condition of Equation \ref{Eq:ZnajekCondition} to select the sign of the toroidal magnetic field, and have used $\Omega(\theta)$ to denote the field line angular velocity as an arbitrary function of poloidal angle.  Far from the black hole those electromagnetic fields may be expressed in standard orthonormal spherical coordinates as:
\begin{align}
\mathbf{B} &= \frac{B_0}{r^2} \hat{r} + \frac{B_0 \Omega(\theta) \sin \theta}{r} \hat{\phi}, \nonumber \\
\mathbf{E} &= -\frac{B_0 \Omega(\theta) \sin \theta}{r} \hat{\theta}.
\end{align}
This field configuration always yields a Poynting flux that is directed radially inward, consistent with the physical demand that energy not be extracted from a non-rotating black hole (enforced by application of the Znajek regularity condition of Equation \ref{Eq:ZnajekCondition}).  As such it might be said that this solution is restricted by conditions at both the horizon and at spatial infinity.  The solutions found by \citet{Michel1973} have uniform rotation, $\Omega(\theta) = \Omega_0$,  and \citet{BZ77} perturbed around the static $\Omega_0 = 0$ solution. 

If $\Omega(\theta) = 1/a \sin^2 \theta$ then the solution of Equation \ref{Eq:SchwarzschildMonopoleSolution} is also valid when applied to an arbitrarily rotating black hole with spin parameter $a$ (\citet{MenonDermer2007} or \citet{GrallaJacobson2014}).  Unfortunately such solutions do not extract a black hole's rotational energy and can become problematic near the poles; we note them here for the sake of completeness but will not address them further.    


\section{Perturbing to Kerr - Most Monopolar} \label{Sec:KerrMostMonopolar}

In this section we arrive at the perturbed monopolar solution of \citet{BZ77} using a slightly different approach in an attempt to elucidate the nature of the solution obtained.  Every solution applicable to a Schwarzchild spacetime found in the previous section (Equation \ref{Eq:SchwarzschildMonopoleSolution}) is ``monopolar'' in the sense that the poloidal magnetic field traces the vacuum poloidal electric field that would be sourced by a non-rotating black hole possessing an electric charge of magnitude $B_0$ (we make this distinction due to magnetic monopoles being on somewhat weaker footing than electric monopoles).  We now determine which solution(s) remain monopolar in that sense when extended to a rotating black hole, using the same general procedure as was used in the previous section.  We begin as before by assuming a monopolar vector potential, except this time in a form appropriate to a rotating spacetime with $a \neq 0$ (which may be derived by taking the standard Kerr-Newman electric field and converting it to a magnetic field):
\begin{equation} \label{Eq:KerrMonopoleVectorPotential}
A_\phi = B_0 \frac{r^2 + a^2}{r^2 + a^2 \cos^2 \theta} \cos \theta.
\end{equation} 
Far from the horizon we arrive at the same condition on the toroidal field as was found in the Schwarzschild case (Equation \ref{Eq:InfinityCondition}, recast in terms of $A_\phi$):
\begin{equation} \label{Eq:InfinityCondition2}
\left[\left(\sqrt{-g} F^{\theta r} \right)^2\right]'_{r \rightarrow \infty} = \frac{1}{B_0^2} \left[(B_0^2 - A_\phi^2)^2 \Omega_\text{F}^2 \right]'_{r \rightarrow \infty}.
\end{equation}
When the spacetime rotates the Znajek regularity condition on the horizon proves to be much more restrictive than it was in the previous section; instead of simply restricting the sign of the toroidal field, we now also require that the toroidal field satisfy (Appendix \ref{App:MostMonopolar}):
\begin{align}
\left(\sqrt{-g} F^{\theta r} \right)^2_\text{H} &= \frac{1}{B_0^2} \left( \Omega_\text{F} - \omega_\text{H} \right)^2 \left(1 - \frac{a^2A_\phi^2}{B_0^2 m^2}\right)_\text{H} \nonumber \\
&\cdot \left[\frac{m^2 B_0^2}{r_\text{H}^2} \left(1 + \sqrt{1 - \frac{a^2A_\phi^2}{B_0^2 m^2}} \right)^2 - A_\phi^2 \right]_\text{H}^2.
\end{align} 
For a slowly rotating black hole, we may proceed by considering an expansion in black hole spin to find:
\begin{align}
\left(\sqrt{-g} F^{\theta r}\right)^2_\text{H} &=  \frac{1}{B_0^2} \left( \Omega_\text{F} - \omega_\text{H} \right)^2 \left[ \vphantom{\mathcal{O} \left(\frac{a^4}{m^4} \right)} \left(B_0^2 - A_\phi^2 \right)^2 \right. \nonumber \\
&\left. + \frac{1}{B_0^2}\left(B_0^2 - A_\phi^2 \right)^3 \frac{a^2}{m^2} + \mathcal{O} \left(\frac{a^4}{m^4} \right) \right]_\text{H}.
\end{align}
Here we have assumed that $\Omega_\text{F} \sim \mathcal{O}(\omega_\text{H})$; if they were of significantly different order then the inner and outer light surfaces would not be distinct and the solution would have diminished physical relevance (assuming black hole spin  different enough from $a = 0$ to be interesting).  Combining this with the condition at spatial infinity (Equation \ref{Eq:InfinityCondition2}) yields:
\begin{equation}
\Omega_\text{F}^2 = \left( \Omega_\text{F} - \omega_\text{H} \right)^2 + \mathcal{O}\left(\frac{a^4}{m^4} \right).
\end{equation}
Therefore for slowly rotating black holes we have $\Omega_\text{F} \approx \omega_\text{H}/2$, the same conclusion \citet{BZ77} and others have arrived at using slightly different approaches.  Dropping the $a^4$ terms to arrive at that result effectively reduced the vector potential of Equation \ref{Eq:KerrMonopoleVectorPotential} to $A_\phi = B_0 \cos \theta$, the same as in the Schwarzschild case.  Should higher order corrections be desired to explore the more strictly ``monopolar'' form of Equation \ref{Eq:KerrMonopoleVectorPotential} or other effects, the full expression for $\Omega_\text{F}$ is given by:
\begin{equation} \label{Eq:OmegaFPureMonopole}
\Omega_\text{F} = \omega_\text{H} \frac{\chi}{\chi + \left(B_0^2  - A_\phi^2\right)},
\end{equation}
where:
\begin{equation}
\chi = \sqrt{1 - \frac{a^2A_\phi^2}{B_0^2 m^2}}\left[\frac{m^2 B_0^2}{r_\text{H}^2} \left(1 + \sqrt{1 - \frac{a^2A_\phi^2}{B_0^2 m^2}} \right)^2 - A_\phi^2 \right].
\end{equation}
Taking an expansion in spin on the horizon, the field line angular velocity is then given by:
\begin{equation} \label{Eq:OmegaFMonopoleExpansion}
\Omega_\text{F}(r_\text{H}) = \omega_\text{H} \left[\frac{1}{2} + \frac{a^2 \sin^2 \theta}{8 m^2} + \frac{3 a^4 \sin^2 \theta}{64m^4} + \mathcal{O}\left(\frac{a^6}{m^6} \right) \right].
\end{equation}
The reason for selecting the techniques used in this section was to note that the assumption of a specific poloidal magnetic field configuration followed by solving for compatible distributions of field line angular velocity and toroidal field is exactly what most perturbation techniques do (at least to leading order).  This is due to the fact that if there were significant changes to the poloidal field a perturbative approach would likely be ill-behaved, as the non-linear nature of the equations involved (most prominently sourced by field line rotation) could easily amplify even small changes to the fields and lead to higher-order corrections that are comparable in magnitude (if not larger) than the unperturbed terms.  

At higher orders the solution obtained here differs slightly from the solutions obtained using other perturbative approaches, although direct comparisons can become difficult: at higher orders other approaches often sacrifice the rigid conservation of field-aligned fluxes of energy and angular momentum and can have difficulty elegantly describing the region interior to $r = 2m$ (i.e. the ergoregion).  Nonetheless such solutions are still broadly compatible with ours, finding $A_\phi \sim \cos \theta + a^2 \cos \theta \sin^2 \theta$ (compatible with the assumed monopolar geometry in Equation \ref{Eq:KerrMonopoleVectorPotential}) and a field line angular velocity $\Omega_\text{F} \sim \omega_\text{H}/2 + a^3 \sin^2 \theta$ (compatible with the expansion in Equation \ref{Eq:OmegaFMonopoleExpansion}).  The differences are primarily due to our approach of exactly satisfying the force-free condition of Equation \ref{Eq:ForceFreeCondition} on only the horizon and at spatial infinity, while perturbative approaches typically seek to approximately satisfy Equation \ref{Eq:ForceFreeCondition} over all space (or at least between the perturbed horizon and spatial infinity).  In practice that means that our solution has error somewhat off the horizon near the equatorial region (Appendix \ref{App:MostMonopolar}), while perturbed solutions have error of similar magnitude but concentrated on and near the horizon. 

In summary, when the spacetime is non-rotating a ``monopolar'' field may have an arbitrary rotational profile described by $\Omega_\text{F} = \Omega(\theta)$ (Section \ref{Sec:SchwarzschildMonopole}).  In this section we found that the solution that most smoothly transitions to a rotating black hole while remaining largely ``monopolar'' is the solution with a rotational profile given by $\Omega(\theta) \approx 0.5 \omega_\text{H}$.  In the next section we will explore how solutions with $\Omega(\theta) \neq 0.5 \omega_\text{H}$ might behave when extended to rotating spacetimes.     

\begin{figure*}[ht]
     \includegraphics[width=\textwidth,clip=true]{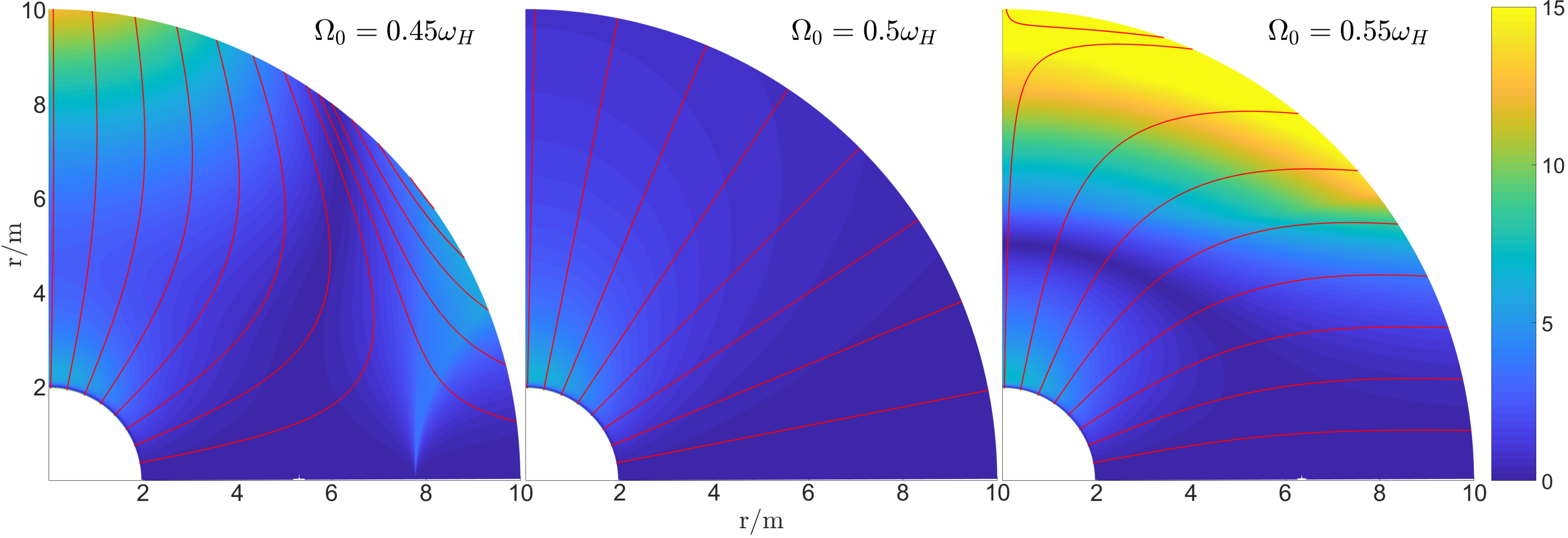}
     \caption{Three magnetospheres with $\Omega_0 = 0.45 \omega_\text{Hp}$, $\Omega_0 = 0.5 \omega_\text{Hp}$, $\Omega_0 = 0.55 \omega_\text{Hp}$ (where $\omega_\text{Hp} \equiv a / 4m^2$) for black hole spin $a = 0.3m$.  The background shading is the percent error of the solutions (\citet{TTT2017}).  The $\Omega_0 = 0.5 \omega_\text{H}$ solution is a separatrix between two classes of solutions that can exhibit significant modifications to the structure of the poloidal field when extended from Schwarzschild to Kerr spacetimes.  We have deliberately chosen to extend the domain to include topological changes to the field (i.e. divergences from monopolarity).  In practice those regions are mostly indicative of a breakdown in the solution, and should be viewed with some suspicion.}  
	   \label{Fig:Figure1}
\end{figure*}


\section{Perturbing to Kerr - Generic Behavior} \label{Sec:KerrGenericExtension}

In the previous section we found that the monopolar magnetosphere in Schwarzschild spacetimes that most closely corresponds to a monopolar magnetosphere in Kerr spacetimes (in the limit of low black hole spin) is specified by $\Omega_\text{F} = \omega_\text{H}/2$.  We now explore what might happen to some of the other $\Omega_\text{F} \neq \omega_\text{H}/2$ solutions from Equation \ref{Eq:SchwarzschildMonopoleSolution} when they are extended to slowly rotating spacetimes.  We note from the outset that the results of the previous section indicate that those extensions might generally be expected to involve significant changes to the structure of the poloidal field; they are unlikely to remain ``monopolar'' over all space, and as such might have only limited regions where the concept of a small perturbation remains valid.  

For simplicity we will only consider uniform (constant) field line angular velocities, expressed as $\Omega_\text{F} = x \omega_\text{H}$, with $x$ a unitless weighting factor on the angular velocity of the horizon. 

We begin by perturbing the metric in spin,\footnote{The results of this section may also be obtained without this conceit through direct perturbation of  the force-free condition of Equation \ref{Eq:ForceFreeCondition} (Appendix \ref{App:ExplicitPerturbation}).  We choose a more circuitous route here and expand the metric and fields separately because doing so can lead to more illuminating intermediate expressions where the different sources of rotational effects are more easily separated and identified.} such that:
\begin{equation}
ds^2 = \left(1 - \frac{2m}{r} \right) dt^2 + \frac{4ma \sin^2 \theta}{r} dt d\phi - \frac{r}{r - 2m} dr^2 - d\Omega^2.
\end{equation} 
Here $d\Omega^2 = r^2 d \theta^2 + r^2 \sin^2 \theta d\phi^2$ and the metric is equivalent to Schwarzschild with the exception of an additional $g_{t \phi}$ component.  We then assume that the monopolar solutions' vector potential gains an additional term when extended to that slowly rotating spacetime:
\begin{equation}
A_\phi = B_0 \cos \theta + a^2 B_0 R(r) \Theta (\theta).
\end{equation}
We demand an $a^2$ correction here because (as noted in the previous section) the field line angular velocity is likely to be proportional  to the horizon's angular velocity if both the black hole spin and magnetosphere rotation are physically interesting, and as signed quantities the two sources of rotation should couple as $a \Omega_\text{F} \sim a^2/m^2$.  

We next demand that the vector potential near the horizon remain unchanged, in the sense that the correction to the vector potential must have $R(2m) = 0$.  The primary reason for this demand is that a generally monopolar field close to the horizon is typically more physically plausible than monopolar behavior further away.  As we expect to find potentially large deviations from monopolarity in some regions for $\Omega_\text{F} \neq \omega_\text{H}/2$ magnetospheres, we choose to directly enforce small changes to the vector potential compatible with our perturbative approach in the more relevant near-horizon region.

Next we determine the structure of the toroidal field; as $R(2m) = 0$, the appropriate form the of horizon regularity condition is straightforward to evaluate to find (in terms of the unitless weighting factor $x$)\footnote{This expression assumes that $\rho_\omega^2 \equiv g_{t \phi}^2 - g_{tt} g_{\phi \phi}$ vanishes when $r = r_\text{H} = 2m$.  For the perturbed metric in use we actually have $\rho_\omega^2 (r_\text{H}) = a^2 \sin^4 \theta$, so this should be understood as a perturbed expression.}:
\begin{align} \label{Eq:GenericToroidal}
\left(\sqrt{-g} F^{\theta r}\right)^2_\text{H} &= \left. \frac{g_{\phi \phi}}{g_{\theta \theta}} \left(\Omega_\text{F} - \omega_\text{H} \right)^2 A_{\phi, \theta}^2 \right|_{\text{H}} \nonumber \\
&= \left.\frac{\omega_\text{H}^2}{B_0^2} \left(x - 1\right)^2 \left(B_0^2 - A_\phi^2 \right)^2\right|_{\text{H}}.
\end{align}
The only remaining task is to evaluate the unknown functions $R(r)$ and $\Theta (\theta)$ in the vector potential.  We do so by taking $\omega_\text{H} = a/4m^2$, then considering the appropriate form of the force-free condition 
of Equation \ref{Eq:ForceFreeCondition}.  We then find that to $\mathcal{O}(a^4/m^4)$ we must have $\Theta(\theta) = \sin \theta \cos^2 \theta$, and demand that $R(r)$ satisfy (primes denoting derivatives with respect to $r$):
\begin{align} \label{Eq:KerrDiffEq}
&8m^4 r \left(r - 2 m\right)^2 R'' + 16 m^5 \left(r - 2m \right) R' \nonumber \\
&- 48 m^4 \left(r - 2m \right) R = r^3 \left(1 - 2x\right) + 16m^3 x.
\end{align}
Re-writing this in terms of the variable $y = r - 2m$ and the unknown function $Y(y)$, we find (primes denoting derivatives with respect to $y$):
\begin{align} \label{Eq:KerrDiffEqY}
&8 m^4 y\left(y + 2m \right) Y'' + 16 m^5 Y' \nonumber \\
&- 48 m^4 Y = \left(1 - 2x \right) \left(y^2 + 6my + 12m^2 \right) + \frac{8 m^3}{y}.
\end{align}
The $8 m^3/y$ term on the right hand side sources undesirable logarithmic terms, so we drop it (dropping this term is also advantageous when excess momentum flux is considered, as discussed in the next section).  Demanding that $Y(0) = R(r_\text{H}) = 0$ we find a general solution of:
\begin{align} \label{Eq:KerrGenericSolution}
A_\phi &= B_0 \cos \theta - \frac{a^2}{16 m^4} \frac{B_0}{9 m} \left(x - \frac{1}{2}\right) \left(r - 2m \right) \nonumber \\
&\cdot \left(44 r^2 + 13 m r + 14 m^2 \right) \cos \theta \sin^2 \theta, \nonumber \\
\Omega_\text{F} &= x \omega_\text{Hp}, \nonumber \\
\sqrt{-g} F^{\theta r}  &= \frac{1}{B_0} \left(x \omega_\text{Hp} - \frac{a}{4m^2}\right) \left(B_0^2 - A_\phi^2 \right).
\end{align}
Here $\omega_\text{Hp} = a/4m^2$; although it has a correspondence to $\omega_\text{H}$, $\omega_\text{Hp}$ should not be taken to vanish as $a \rightarrow 0$; rather it should be taken as a convenient description of $\Omega_\text{F}$ once an $a \neq 0$ spacetime has been specified.  In application the above solution should be applied with both the metric and force free condition of Equation \ref{Eq:ForceFreeCondition} in their full form ($\omega_\text{Hp}$ should remain as $a/4m^2$); the separate perturbed expressions shown in this section are only used to explain the considerations used to arrive at Equation \ref{Eq:KerrGenericSolution}.  

The solution of Equation \ref{Eq:KerrGenericSolution} makes it clear that the most monopolar $x = 1/2$ solution found in the previous section is a separatrix between two different types of solutions.  It also makes it clear that the perturbative approach used can become problematic; the correction to the poloidal magnetic field is unbounded as $r \rightarrow \infty$ unless $x = 1/2$.  We discuss those outcomes and the error of the solution in the next section.


\section{Error Analysis and Behaviors} \label{Sec:ErrorAnalysis}

In this section we discuss the error and general behaviors of the perturbative extension of arbitrarily uniformly rotating magnetospheres found in the previous section, the role of the outer boundary surface, and the bending behaviors found.  

\subsection{General Solution Error}

The general solution found in the previous section (Equation \ref{Eq:KerrGenericSolution}) yields an excess momentum flux (Equation \ref{Eq:ForceFreeCondition}) that is given by ($\Omega_\text{F} = x \omega_\text{Hp}$):
\begin{equation} \label{Eq:ExcessMomentumFlux}
4 \pi \Sigma \frac{X_A}{A_{\phi, A}} =  a^2 B_0 \frac{2m + r}{m r^4} \cos\theta + \mathcal{O} \left(\frac{a^4}{m^4} \right).
\end{equation}
This form is a reason for dropping the $8 m^3/y$ term when solving for $Y(y)$ in Equation \ref{Eq:KerrDiffEqY}; to order $a^2$ the failure of the momentum flux $X_A$ to vanish is then identical for all magnetospheres (including the standard ``first order'' perturbed monopole solution of \citet{BZ77}).  The $\mathcal{O}(a^4)$ term goes as $r^3$ (as a function of $x$), so although the excess momentum flux $X_A$ is of comparable magnitude for all solutions near the horizon, it generally grows with increasing $r$ for $x \neq 1/2$ as the vector potential begins to deviate ever more strongly from its $a \rightarrow 0$ monopolar behavior.    

A representation of that error is shown in Figure \ref{Fig:Figure1} for three different $x$ values and black hole spin $a = 0.3m$.  The error is expressed as the full excess momentum flux $X_A$ related as a percentage of the largest term of the force-free condition in Equation \ref{Eq:ForceFreeCondition} (the exact method used may be found in \citet{TTT2017}).  It is apparent that the percent error of $x \neq 1/2$ solutions grows in $r$.  As was shown in Section \ref{Sec:KerrMostMonopolar}, that is because $x = 1/2$ is the solution most compatible with monopolarity at both the horizon and spatial infinity.

It is also apparent that the topology of the poloidal magnetic field lines changes for $x \neq 1/2$.  The field lines that do not intersect the horizon are outside the domain of the toroidal field as a function of vector potential found in Equation \ref{Eq:GenericToroidal}, however, so the exact nature of those topological changes should be viewed with some suspicion.  For the purposes of this work we do not consider such topological changes to be anything more than an indication that the perturbation techniques might have been extended outside a domain of more robust validity, and we don't insist upon their existence.  Should a more extensive radial domain be desired we would suggest either consideration of solutions closer to $x = 1/2$ or the application of other solution techniques before any attempt to justify such topological changes. 

The more robust structural change is the tendency for the poloidal magnetic field lines to begin bending away from monopolarity for $x \neq 1/2$.  In Figure \ref{Fig:Figure1} we have chosen solutions that exhibit fairly significant bending close to the black hole; if for whatever reason an outer boundary further from the black hole were desired, those solutions might be problematic.  Nonetheless, for any boundary surface located a finite distance from the black hole there will be a range of field line angular velocities described by $x = 1/2 \pm \epsilon$ that will exhibit the same bending tendencies while maintaining error comparable to the error of the standard $x = 1/2$ solution.  

\subsection{Domain Boundary Surfaces}

In solving for the force-free magnetospheres around non-rotating black holes in Section \ref{Sec:SchwarzschildMonopole} and the most monopolar solution in Section \ref{Sec:KerrMostMonopolar} we applied conditions at both spatial infinity and the horizon (though spatial infinity was a mathematical convenience, not a requirement; a boundary surface located at finite radii would lead to the same overall conclusions).  The general solution found in Section \ref{Sec:KerrGenericExtension} did not apply any conditions at spatial infinity, however, and as such can have potentially significant error for some $x$ values at larger radii, which might lead to that solution being viewed as less robust or more erroneous.

That might be true in some instances, but it should be noted what the solution procedures applied in Sections \ref{Sec:SchwarzschildMonopole} and \ref{Sec:KerrMostMonopolar} are actually doing: they're demanding a specific poloidal field geometry over all space, then finding conserved fluxes of energy and angular momentum compatible with that assumption.  While that is a mathematically useful approach, it tacitly allows the conditions on an (at least) super-Alfv\'{e}nic plasma inflow close to the horizon to directly feed back and communicate with a plasma outflow extended to spatial infinity.

There is also a more fundamental problem in extending the outer boundary to spatial infinity: such magnetospheres fairly generically contain infinite amounts of energy and are not physically realizable.  Significantly restricting the conditions on near horizon behaviors using spatial infinity should therefore be viewed as a potentially useful mathematical technique or simplification, with the knowledge that $r \rightarrow \infty$ is only a stand-in for a more appropriate outer boundary.  

Mathematically that boundary might be taken as $r \gg m$ or similar, but physical considerations are potentially more restrictive.  One such consideration, as suggested above, as that the ingoing near horizon magnetosphere might be expected to be at least somewhat independent of the more distant outgoing magnetosphere.  The simplest example of such a magnetosphere would be a single spherical surface serving as an outer boundary for an inflow and inner boundary for an outflow, but that is not guaranteed.  More generally (from a perspective of physical modeling) one might have an ``inner magnetosphere'' close to the horizon that is connected to a series of different outer magnetosphere regions (described by differing physical approximations and/or variables) that only loosely connect the near horizon region with spatial infinity (or $r \gg m$).  

In \citet{BZ77} ``spark gaps'' and other mechanisms between the inner and outer light surfaces are postulated and discussed specifically because conditions on the horizon and distant regions are intrinsically incompatible without some kind of intermediate joining mechanism or structure where additional physics (beyond the rigid application of stationary and axisymmetric force-free magnetohydrodynamics) must be considered.  The magnitude of any resultant decoupling of inner and outer magnetospheres is an open question that will necessarily vary from model to model and magnetosphere to magnetosphere, but it is clear that in general the horizon and distant regions cannot be self-consistently connected by a single, unbroken magnetic field line described solely by the core assumptions used here and in \citet{BZ77}.  

Depending upon the problem being explored, a practical outer boundary for the inner magnetosphere might lie somewhere interior to the outer light surface (\citet{MacDonaldThorne1982}), perhaps near the separation surface (\citet{TNTT90}).  Regardless of the selection made, however, it is in general somewhat implausible to expect to directly drive near horizon magnetosphere behaviors using a single fixed model rigidly extended to spatial infinity (and vice-versa).  Models that do rigidly connect both regions can nonetheless still be useful, such as in demonstrating that the single physical assumption of a force-free magnetosphere can in principle transmit energy from the horizon to spatial infinity.  However, in a more general exploration of energy-extracting black hole magnetospheres it is overly restrictive to demand a rigid connection between the near horizon inflowing magnetosphere and an outflowing magnetosphere extended to spatial infinity.

\subsection{Bending Behaviors}

For $x < 1/2$ (where $\Omega_\text{F} = x \omega_\text{Hp}$) we found that field lines bend upwards towards the azimuthal axis; for $x > 1/2$ we found that field lines bend downwards towards the equatorial plane (compatible with the numerical results of \citep{TTT2017}).  That behavior is independent of any (finite) outer boundary selected, although the range of reasonable $x$ values becomes more restricted as the outer boundary moves radially outwards.   

No matter what outer boundary is selected, however, if the boundary condition along the equatorial plane (or other ``straight'' boundary) is close to being ``monopolar'' (i.e. a single magnetic field line tracing the boundary) then the presence of upward bending field lines in force-free magnetospheres should not be surprising wherever $\Omega_\text{F} \lesssim 0.5 \omega_\text{H}$ and the presence of downward bending field lines should not be surprising wherever $\Omega_\text{F} \gtrsim 0.5 \omega_\text{H}$.  The primary question is the exact nature of the bending. 

The magnitude of the bending should be expected to increase with increases in black hole spin, distance from the horizon, and in deviations from $\Omega_\text{F} \sim 0.5 \omega_\text{H}$, as such changes will change the strength of the toroidal field and/or the distance over which the toroidal field can force the poloidal field to bend.  Numerical experiments in \citet{TTT2017} confirm those tendencies.  However we would hesitate to call such tendencies anything more than a potentially useful ``rule of thumb''; additional complications, such as the simultaneous presence of opposing tendencies (i.e. regions containing both $x < 1/2$ and $x > 1/2$) or the presence of different boundary conditions could break the ``rule''.

We are not the first to suggest a general rule coupling magnetosphere bending behaviors to field line rotation.  In \citet{Penna2015} impedance matching arguments were made using resistive membranes on the horizon and at spatial infinity (to include surfaces approximating spatial infinity) to suggest that field lines with diverging angular separation should have $\Omega_\text{F} \gtrsim 0.5 \omega_\text{H}$ and that field lines with converging angular separation should have $\Omega_\text{F} \lesssim 0.5 \omega_\text{H}$.  In other words they also expect $\Omega_\text{F} \sim 0.5 \omega_\text{H}$ to generally be a separatrix between magnetosphere bending behaviors, with a monopolar configuration coinciding with the separatrix.

The ability to adjust the bending of magnetic field lines by adjusting field line angular velocity can add both significant flexibility and restrictions when considering black hole energy extraction.  The single ``mostly monopolar'' solution originally obtained by \citep{BZ77} provides only a single rate of energy extraction and angular momentum outflow in a single fixed direction.  By changing field line angular velocity the rates of energy and angular momentum extraction can be easily modified in way that is coupled to the direction of their outflow.  

The direct formation of jets, for example, might be aided by more slowly rotating magnetospheres.  More rapidly rotating magnetospheres, meanwhile, might more easily connect to a nearby accretion disk.  Both behaviors would diminish the overall rates of energy extraction while simultaneously enhancing or reducing the rates of angular momentum extraction, potentially limiting the timescales over which such magnetospheres might be relevant or applicable.


\section{Conclusions}

It was noted in \citet{BZ77} that solving for magnetospheres via a perturbation in black hole spin might be ill-advised unless the poloidal magnetic field remains essentially unchanged.  That is a reason why only a single solution from the general class of monopolar solutions found by \citet{Michel1973} is typically treated in analytic extensions to rotating spacetimes.  Although the more general solution space is more difficult to compute it is still of physical interest, and indicates that slowly rotating magnetospheres might be expected to bend towards the azimuthal axis while more rapidly rotating magnetospheres might be expected to bend towards the equatorial plane.  Attempts to refine the analytic treatment of the ``mostly monopolar'' $\Omega_\text{F} \approx 0.5 \omega_\text{H}$ solution might also benefit from the knowledge that it is a separatrix between two classes of behaviors, as initially ignorable deviations from the separatrix might lead to significant effects as the solution is refined.

The ``mostly monopolar'' solution has value in that it simultaneously provides both an inflow near-horizon solution and an outflow solution that can be extended to spatial infinity.  However, there is no requirement that a near-horizon inflow solution be rigidly coupled to an outflow solution.  In fact almost the complete opposite is true, and for internal self-consistency inflow and outflow solutions must be decoupled to at least some extent.  Disregarding inflow solutions solely because they do not extend to spatial infinity therefore artificially limits understanding of black hole energy extraction.  There can be good reasons for desiring such an extension, but applying such a restriction necessarily limits any solutions obtained to special cases, just as the ``mostly monopolar'' solution is a special case of a more general solution space.


\section*{Acknowledgments} 

M.T. was supported by JSPS KAKENHI Grant Number 17K05439, and DAIKO FOUNDATION.


%


\begin{appendix}

\begin{figure}[ht]
     \includegraphics[width=\columnwidth,clip=true]{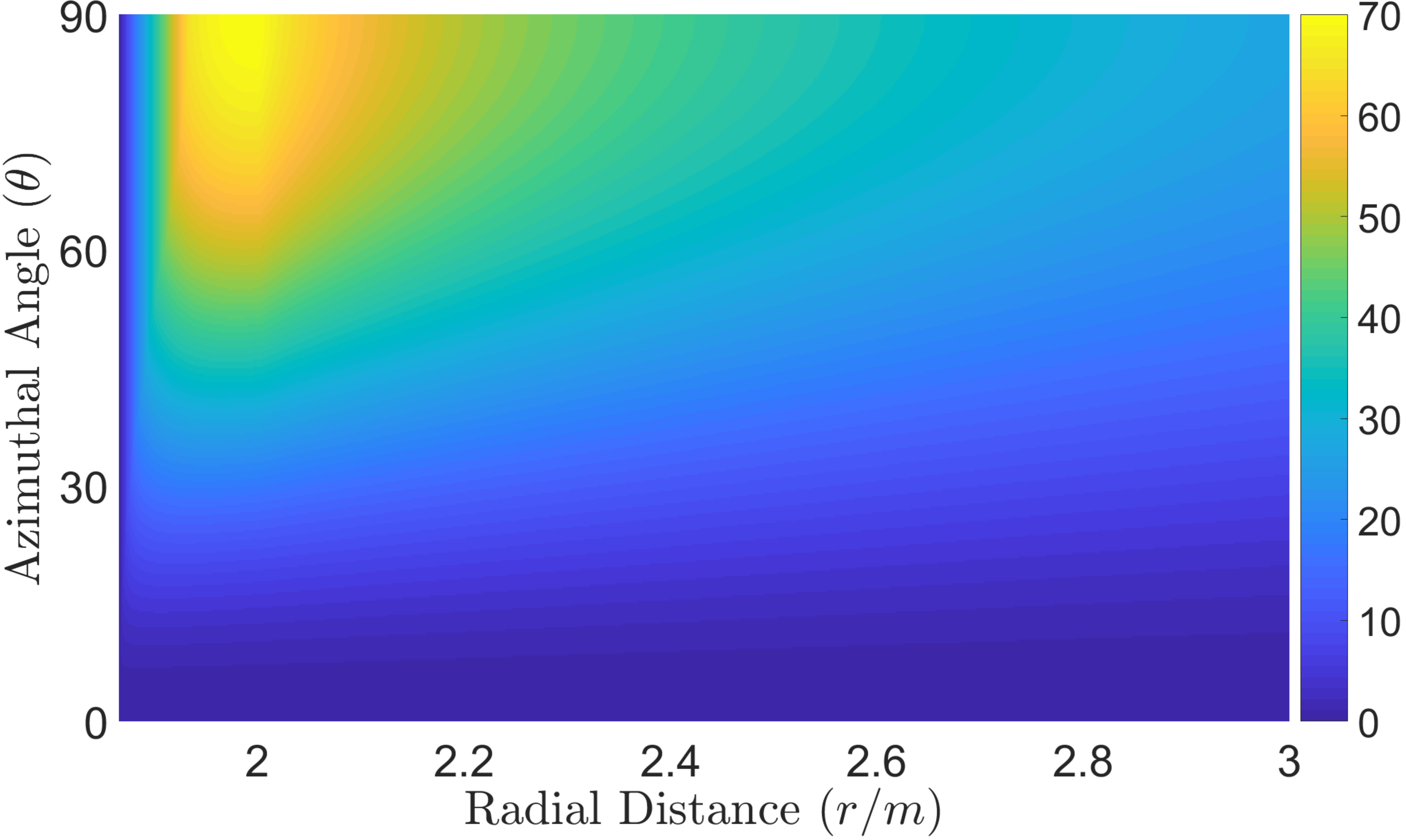}
     \caption{The percent error (\citet{TTT2017}) of a monopolar vector potential (Appendix \ref{Eq:AppKerrMonopole}, Section \ref{Sec:KerrMostMonopolar}) with black hole spin parameter $a = 0.5m$. The error is largest near the equatorial plane around $r=2m$, but vanishes along the horizon and as $r \rightarrow \infty$.  The increase near the equatorial plane is a result of the field line angular velocity there being too large ($\Omega_\text{F} \gtrsim 0.5 \omega_\text{H}$) to be completely compatible with a ``straight'' magnetic field line.  Standard perturbed monopole solutions have errors of similar magnitude, but concentrated on and near the horizon.} 
	   \label{Fig:Figure2}
\end{figure}

\section{Explicit Monopole and a Rotating Black Hole} \label{App:MostMonopolar}

The toroidal field should be a function of the vector potential $A_\phi$ due to its correspondence with a conserved flux of angular momentum from the assumption of axisymmetry.  In this Appendix we solve for that function in detail if the poloidal magnetic field on the horizon is specified as an explicit monopole.  A monopolar vector potential is given by (Equation \ref{Eq:KerrMonopoleVectorPotential}):
\begin{equation} \label{Eq:AppKerrMonopole}
A_\phi = B_0 \frac{r^2 + a^2}{\Sigma} \cos \theta.
\end{equation}	
Here $\Sigma = r^2 + a^2 \cos^2 \theta$, and with $A_t = - B_0 a \cos \theta / \Sigma$ the vector potential would describe the vacuum solution of a black hole possessing a magnetic charge of magnitude $B_0$.  On the horizon the toroidal magnetic field is given by (Equation \ref{Eq:ZnajekCondition}):  
\begin{align}
\sqrt{-g} F^{\theta r}_\text{H} &= -\frac{\left(r_\text{H}^2 + a^2 \right) \left(\Omega_\text{F} - \omega_\text{H} \right) \sin \theta}{\Sigma_\text{H}} A_{\phi_\text{H}, \theta} \nonumber \\
&= -\left(\Omega_\text{F} - \omega_\text{H} \right) \frac{A_\phi \tan \theta}{B_0} A_{\phi_\text{H}, \theta}.
\end{align}
In order to express this purely in terms of $A_\phi$ we first evaluate $A_{\phi, \theta}$ to find:
\begin{align}
A_{\phi, \theta} &= B_0 \left(r^2 + a^2 \right) \left(\frac{2a^2}{\Sigma^2} \sin \theta \cos^2 \theta - \frac{1}{\Sigma} \sin \theta \right) \nonumber \\
&= A_\phi \left(1 - \frac{2r^2}{\Sigma} \right) \tan \theta.
\end{align}
Therefore the toroidal field on the horizon is given by:
\begin{align}
\sqrt{-g} F^{\theta r}_\text{H} &= -\frac{1}{B_0}\left(\Omega_\text{F} - \omega_\text{H} \right) \left(1 - \frac{2r^2_\text{H}}{\Sigma_\text{H}} \right) A_{\phi_\text{H}}^2 \tan^2 \theta \nonumber \\
&= -\frac{1}{B_0}\left(\Omega_\text{F} - \omega_\text{H} \right) \left(1 - \frac{2r^2_\text{H}}{\Sigma_\text{H}} \right) \nonumber \\
&\cdot\left[\left(\frac{2 m r_\text{H} B_0}{\Sigma_\text{H}}\right)^2 - A_{\phi_\text{H}}^2 \right].
\end{align}
To proceed further we require an expression for $\Sigma(A_\phi)$ on the horizon; from the expression for $A_\phi$ given in Equation \ref{Eq:AppKerrMonopole} we find:
\begin{equation}
A_\phi^2 = \frac{B_0^2}{\Sigma^2}\left(r^2 + a^2 \right)^2 \frac{\Sigma - r^2}{a^2}.
\end{equation}  
Solving for $1/\Sigma$, we find:
\begin{equation}
\frac{1}{\Sigma} = \frac{1}{2r^2} \left(1 \pm \sqrt{1 - 4 \frac{a^2 r^2 A_\phi^2}{B_0^2 \left( r^2 + a^2\right)^2}} \right).
\end{equation}
Taking the relevant positive branch and evaluating along the horizon, we find:
\begin{equation}
\frac{2 r_\text{H}^2}{\Sigma_\text{H}} = 1 + \sqrt{1 - \left(\frac{a A_{\phi_\text{H}}}{B_0 m}\right)^2}.
\end{equation}
We insert this expression into the horizon condition on the toroidal field to conclude that:
\begin{align}
\sqrt{-g} F^{\theta r}_\text{H} &= \frac{1}{B_0}\left(\Omega_\text{F} - \omega_\text{H} \right) \sqrt{1 - \left(\frac{a A_{\phi_\text{H}}}{B_0 m}\right)^2} \nonumber \\
&\cdot\left[\frac{m^2 B_0^2}{r_\text{H}^2} \left(1 + \sqrt{1 - \frac{a^2A_{\phi_\text{H}}^2}{B_0^2 m^2}} \right)^2 - A_{\phi_\text{H}}^2 \right].
\end{align}
In Section \ref{Sec:KerrMostMonopolar} this is used to calculate the field line angular velocity (Equation \ref{Eq:OmegaFPureMonopole}) for a monopole by comparing the above condition to the condition on the toroidal field at spatial infinity.  The resultant field line angular velocity on the horizon goes as $\Omega_\text{F}(r_\text{H}) \sim \omega_\text{H} (0.5 + \sin^2 \theta)$ (Equation \ref{Eq:OmegaFMonopoleExpansion}).  The relatively significant deviation away from $\Omega_\text{F} \approx 0.5 \omega_\text{H}$ near the equatorial plane is generally incompatible with a ``straight'' magnetic field line in the poloidal plane, leading to error near the equatorial plane for median values of $r$ (Figure \ref{Fig:Figure2}).  The first step towards eliminating such error might be to explicitly add more ``bunching'' of field lines near the azimuthal axis; such bunching is observed when finding solutions numerically (eg. \citep{TTT2017}), and is discussed in more detail in \citet{GrallaEtAl2015}. 

\section{Alternative Perturbation} \label{App:ExplicitPerturbation}

In this Appendix we solve for the generic perturbed solution found in Section \ref{Sec:KerrGenericExtension} by more directly considering the force-free transfield equation (Equation \ref{Eq:ForceFreeCondition}).  We first note that the most monopolar solution of Section \ref{Sec:KerrMostMonopolar} may be expanded in spin (as originally arrived at by \citet{BZ77}) to find:
\begin{align}
A_\phi &= B_0 \cos \theta, \nonumber \\
\Omega_\text{F} &= \frac{1}{2} \frac{a}{4 m^2}, \nonumber \\
\sqrt{-g} F^{\theta r} &= -B_0 \frac{a}{8 m^2} \sin^2 \theta \nonumber \\
&= -\frac{1}{B_0} \frac{a}{8 m^2} \left(B_0^2 - A_\phi^2 \right).
\end{align}
In other words (at least to leading order) the reaction of the initially non-rotating magnetosphere to the addition of spacetime rotation is to develop outward fluxes of energy and angular momentum while maintaining the same poloidal magnetic field structure.  This solution may be placed into the force-free transfield equation to find its corresponding excess momentum flux (defining $\bar{X}$ as $\bar{X} \equiv 4 \pi \Sigma \sin \theta X_A / A_{\phi, A}$):
\begin{align}
\bar{X} &= -\frac{1}{2} \frac{\Sigma}{\Delta \sin \theta} \frac{d}{d A_\phi} \left( \sqrt{-g} F^{\theta r}\right)^2 - \frac{1}{\Delta} \left(\frac{\alpha}{\sin \theta} A_{\phi, \theta} \right)_{, \theta} \nonumber \\
&= \frac{2 B_0 \Sigma}{\Delta} \left(\frac{a}{8 m^2} \right)^2 \cos \theta \sin \theta \nonumber \\
&+ \frac{B_0}{\Delta} \left[g_{tt, \theta} + 2 g_{t \phi, \theta} \left(\frac{a}{8 m^2}\right) + g_{\phi \phi, \theta} \left(\frac{a}{8 m^2}\right)^2 \right].
\end{align}
The division by zero divergence of this expression on the horizon ($\Delta = 0$) is the result of failing to satisfy the square of the Znajek regularity condition of Equation \ref{Eq:ZnajekCondition}, which is intrinsic to the transfield equation.  The metric elements $g_{tt}$, $g_{t \phi}$, and $g_{\phi \phi}$ may be expanded in spin to find:
\begin{align}
g_{tt} &= 1 - \frac{2m}{r} + \frac{2m \cos^2 \theta}{r^3} a^2 +  \mathcal{O}\left(a^4 \right), \nonumber \\
g_{t \phi} &= \frac{2m \sin^2 \theta}{r} a + \mathcal{O}\left(a^3 \right), \nonumber \\
g_{\phi \phi} &= -r^2 \sin^2 \theta - \frac{r + 2m \sin^2 \theta}{r} a^2 \sin^2 \theta  +  \mathcal{O}\left(a^4 \right).  
\end{align}  
If we discard the metric elements of order $a^2$ and higher (taking $\Sigma = r^2$), the excess momentum flux becomes:
\begin{align}
\bar{X} &\approx \frac{2 B_0 r^2}{\Delta} \left(\frac{a}{8 m^2} \right)^2 \cos \theta \sin \theta \nonumber \\
&+ \frac{B_0}{\Delta} \left[\frac{8 m a}{r}\left(\frac{a}{8 m^2}\right) - 2 r^2 \left(\frac{a}{8 m^2}\right)^2 \right] \cos \theta  \sin \theta  \nonumber \\
&= \frac{B_0}{\Delta} \frac{a^2}{mr} \cos \theta \sin \theta.
\end{align}
This excess momentum flux is due entirely to the $g_{t \phi, \theta} \Omega_\text{F}$ term.  If we also include the order $a^2$ term in $g_{tt}$, we arrive at (taking $\Delta = r^2 - 2mr$):
\begin{align} \label{Eq:AppendixError}
\bar{X} &= \frac{B_0}{\Delta} \left[-\frac{4m a^2}{r^3} \cos \theta \sin \theta + \frac{a^2}{mr} \cos \theta \sin \theta \right] + \mathcal{O}\left(a^4\right) \nonumber \\
&= B_0 \frac{2m + r}{m r^4} a^2  \cos \theta \sin \theta + \mathcal{O}\left(a^4\right).
\end{align}
This is compatible with the excess momentum flux reported above when discussing error in Section \ref{Sec:ErrorAnalysis} (Equation \ref{Eq:ExcessMomentumFlux}).  We have belabored the above in order to suggest that perturbative approaches applied to the problem of black hole magnetospheres can have philosophical differences.  For example, the fundamental perturbation might be taken to be the transfield equation, such that a somewhat inconsistent perturbation in the metric and fields (as above, where the second-order correction to $g_{\phi \phi}$ is irrelevant) is acceptable.  Alternatively, the metric might be taken as the fundamental perturbation, with (for example) only terms of order $a^2$ kept in the metric, and the fields adjusted in whatever manner might be useful in order to find a vanishing momentum flux under application of that perturbed metric.        

In this work we largely followed \citet{BZ77} and took the transfield equation to be the fundamental perturbation.  However, when studying the equations involved it can sometimes be more illuminating to consider alternative approaches.  In light of that, we took a hybrid approach in Section \ref{Sec:KerrGenericExtension} when exploring the behavior of arbitrarily rotating magnetospheres.  We first perturbed the metric in spin, then selected the well-behaved electromagnetic fields corresponding to that metric.  When exploring the error of that solution in Section \ref{Sec:ErrorAnalysis}, however, we took the approach used to arrive at Equation \ref{Eq:AppendixError}, which is to say we treated the transfield equation as the fundamental perturbation.

We will now arrive at the same generically rotating solution found in Section \ref{Sec:KerrGenericExtension} without applying a hybrid approach.  We begin by taking the unperturbed fields around a non-rotating black hole to be given by:
\begin{align}
A_\phi &= B_0 \cos \theta, \nonumber \\
\Omega_\text{F} &= x \frac{a}{4 m^2}, \nonumber \\
\sqrt{-g} F^{\theta r} &= \frac{1}{B_0} x \frac{a}{4 m^2} \left(B_0^2 - A_\phi^2 \right).
\end{align}
Here $x$ is a unitless weighting factor and $a$ is arbitrary, though it will be taken to correspond to the spin of the rotating black hole (such that the inner and outer light surfaces can be taken to be distinct for $x$ values of order $1$).  We now wish to solve for the structure of the fields under the constraint that the conserved field line angular velocity along each field line remains the same.  Following identical logic to that applied in Section \ref{Sec:KerrGenericExtension} (which is insensitive to the metric used under the transformation $2m \rightarrow r_\text{H}$) we then assume that leading order corrections to the fields are given by:  
\begin{align}
A_\phi &= B_0 \cos \theta + a^2 B_0 R\left(r\right) \Theta \left(\theta \right), \nonumber \\
\Omega_\text{F} &= x \frac{a}{4 m^2}, \nonumber \\
\sqrt{-g} F^{\theta r} &= \frac{1}{B_0} \left(x \frac{a}{4 m^2} - \frac{a}{4m^2}\right) \left(B_0^2 - A_\phi^2 \right).
\end{align}
Inserting those fields into the transfield equation, we find (keeping only terms of order $a^2$, and taking advantage of the calculations done above for the monopolar case):
\begin{align}
\bar{X} &= -\frac{1}{2} \frac{\Sigma}{\Delta \sin \theta} \frac{d}{d A_\phi} \left(\sqrt{-g} F^{\theta r}\right)^2 \nonumber \\
&- \frac{1}{\sin \theta} \left(\alpha A_{\phi, r} \right)_{, r} - \frac{1}{\Delta} \left(\frac{\alpha}{\sin \theta} A_{\phi, \theta} \right)_{, \theta} \nonumber \\
&\, \nonumber \\
&= \frac{2 B_0 \Sigma}{\Delta} \left(1 - 2x\right) \left( \frac{a}{4 m^2}\right)^2 \cos \theta \sin \theta \nonumber \\
&+ \frac{B_0}{\Delta} \frac{2 x a^2}{mr} \cos \theta \sin \theta - \frac{4 B_0 m a^2}{\Delta r^3} \cos \theta \sin \theta \nonumber \\
&- \frac{B_0}{\sin \theta} \Theta(\theta) a^2 \left[\left(1 - \frac{2m}{r} \right) R'(r) \right]_{, r} \nonumber \\
&- \frac{B_0}{\Delta} R(r) a^2 \left[ \left(1 - \frac{2m}{r} \right) \frac{\Theta'(\theta)}{\sin \theta} \right]_{, \theta}. 
\end{align}
We can now note two things.  First, we would prefer that $\Theta(\theta) \sim \cos \theta \sin^2 \theta$ such that all terms have the same $\theta$ dependence.  Second, the only $a^2$ correction to the metric to survive is the same as was used above, the correction from $g_{tt}$ that is independent of $x$.  If we demand that the error (in terms of the excess momentum flux) of the general solution match the error of the monopolar solution to order $a^2$, then we must have:
\begin{align} \label{Eq:AppendixDiffEQ}
&\left(\frac{r^2}{8 m^4} - \frac{1}{m r} \right) \frac{1 - 2x}{r^2 - 2mr} - \left[\left(1 - \frac{2m}{r} \right) R'(r) \right]_{, r} \nonumber \\
&+ \frac{6 R(r)}{r^2 - 2mr} \left(1 - \frac{2m}{r} \right) = 0.
\end{align}
Making the variable substitution $y = r - 2m$ for a function $Y(y)$, this becomes:
\begin{align}
&\frac{y}{y + 2m} Y''(y) + \frac{2m}{\left(y + 2m \right)^2} Y'(y) - \frac{6}{\left(y + 2m \right)^2} Y(y) \nonumber \\
&-\frac{y^2 + 6m y + 12m^2 }{8m^4} \frac{1 - 2x}{\left(y + 2m \right)^2} = 0.
\end{align} 
This is compatible with Equation \ref{Eq:KerrDiffEqY}, arrived at through different considerations in Section \ref{Sec:KerrGenericExtension}.  The primary difference is that in Section \ref{Sec:KerrGenericExtension} we ignored the order $a^2$ term in $g_{tt}$ and did not add in a term related to $g_{t \phi}$ in order to explicitly force the errors of the solutions to match.  Instead we dropped an undesirable logarithmic source term from the differential equation, ultimately finding that doing so resulted in exactly matching error.

Our basic goal was to find arbitrarily uniformly rotating magnetospheres with error comparable to the widely known first-order corrections to a monopolar magnetic field found by \citet{BZ77} (from the same philosophical perspective of treating the transfield equation as the primary perturbation).  Ultimately, however, we are perturbing magnetospheres that already possess rotation and angular momentum fluxes, so in a general perturbation in black hole spin it is not entirely obvious that any two factors of $a$ are truly comparable (or simultaneously vanish as black hole spin vanishes).  As such, in Section \ref{Sec:KerrGenericExtension} we took a hybrid approach that attempted to emphasize that it might be more appropriate to consider a slowly rotating spacetime as a distinct perturbation, from which appropriate electromagnetic fields should be calculated in whatever manner might be most convenient.

As a final concluding remark, for the sake of completeness we should point out that \citet{BZ77} did include a correction to $A_\phi$ in order to eliminate the remaining order $a^2$ error shown in Equation \ref{Eq:AppendixError}.  That correction is common to our solution (essentially a particular solution of Equation \ref{Eq:AppendixDiffEQ} with $x = 1/2$ and the order $a^2$ error as a source term).  We have suppressed that correction as being uninteresting.  It is sourced by corrections to the metric that are only relevant inside the ergosphere and as such falls off at large radius, which can be seen by taking the limit $r \gg 2m$ in Equation \ref{Eq:AppendixDiffEQ} with the error of Equation \ref{Eq:AppendixError} as a source term:
\begin{equation}
R''(r) - \frac{6}{r^2} R(r) = \frac{1}{mr^3}.
\end{equation}
Solving, one can conclude that $R(r) \sim 1/4r$, which as expected isn't significant outside the ergosphere.  If the full correction is nonetheless desired, it may be written as:
\begin{align}
R_\text{Corr} &= -\frac{1}{m^4} \left[\frac{m^2 + 3mr - 6r^2}{12} \ln \left(\frac{r}{2m} \right) + \frac{11 m^2}{72} \right. \nonumber \\
& \left. \vphantom{\frac{m^2 + 3mr - 6r^2}{12} \ln \left(\frac{r}{2m} \right)} \frac{m^3}{3r} + \frac{mr}{2} - \frac{r^2}{2} \right] \nonumber \\
&- \left(\frac{2r^3 - 3mr^2}{8m^5} \right) \left[\text{Li}_2 \left(\frac{2m}{r}\right) \right. \nonumber \\
&- \left. \vphantom{\text{Li}_2 \left(\frac{2m}{r}\right)} \ln \left(1 - \frac{2m}{r} \right) \ln \left(\frac{r}{2m} \right) \right],
\end{align}  
where $\text{Li}_2$ is the dilogarithm, defined as:
\begin{equation}
\text{Li}_2 (x) = \int_x^0 \frac{1}{t} \ln \left(1 -t \right) dt.
\end{equation}
The addition of the above correction to the vector potential might in some sense be more mathematically correct, but offers no real physical insight into the problem.  We have therefore ignored it, as it would only serve to obfuscate the more fundamental behaviors involved and complicate the application of the solution to spatial regions interior to $r = 2m$.    

\vfill\clearpage

\end{appendix}



\end{document}